\documentclass[twocolumn, pra, showpacs,superscriptaddress]{revtex4}
\usepackage{graphicx}
\usepackage{amsmath}
\usepackage{float}
\usepackage{color}

\begin{document}

\title{Quantum pump for counter-circulations in a spinor Bose-Einstein condensate}
\author{Z. F. Xu}
\affiliation{Department of Physics, Tsinghua University, Beijing
100084, People's Republic of China}
\author{P. Zhang}
\affiliation{ERATO, JST, Macroscopic Quantum Control Project, Hongo,
Bunkyo-Ku, Tokyo 113-8656, Japan}
\author{R. L\"u}
\affiliation{Department of Physics, Tsinghua University, Beijing
100084, People's Republic of China}
\author{L. You}
\affiliation{Department of Physics, Tsinghua University, Beijing
100084, People's Republic of China}

\begin{abstract}
We propose a pump scheme for quantum circulations, including
counter-circulations for superposition states, of a spinor
Bose-Einstein condensate. Our scheme is efficient and can be
implemented within current experimental technologies and setups. It
remains valid for non-classical atomic states, such as pseudo-spin
squeezed states and maximal entangled N-GHZ or NooN states.
Moreover, it is capable of transforming several enhanced sensing
protocols relying on reduced fluctuations from quantum correlation
and entanglement of atomic internal states to enhanced measurement
of spatial interference and rotation.
\end{abstract}



\maketitle

The existence of superposition states is a hallmark of quantum
mechanics. Distinctly nonclassical properties such as quantum
entanglement can arise for superpositions of more than one degree of
freedom. With proper quantification schemes, entanglement,
essentially ubiquitous in multi-party superposition states, is
viewed as an important resource for quantum information processing
and quantum computation. Generation of superposition states with
high entanglement resource is a challenging experimental goal
confronted many technique difficulties, not the least being
decoherence of a superposition from interacting with environment and
reservoirs. Atomic quantum gases, with its demonstrated long
coherence time and controllability under many circumstances, are
ideal candidates for investigating quantum coherence properties
\cite{Leggett2001}.

This Letter proposes a mechanism for generating superposition states
of counter-circulations in a spinor atomic Bose-Einstein condensate
(BEC), analogous to entangled states between atomic spin and its
center of mass orbital rotation \cite{Pu2001}. Recently, making use
of the orbital angular momentum (OAM) of light, a controllable
scheme was proposed \cite{Kapale2005,Thanvanthri2008}, capable of
creating an arbitrary superposition of vortices and anti-vortices.

Generating vortices in a condensate using OAM from a
Laguerre-Gaussian (LG) light beam was suggested long ago
\cite{Marzlin1997,Bolda1998,Dum1998,Nandi2004,Kanamoto2007}, and
confirmed experimentally in various Raman coupling schemes
\cite{Andersen2006,ryu2007,Wright2008,Wright2009}. An alternative,
some times termed a vortex pump idea, is capable of continuously
changing vorticity through repeated manipulations of external
magnetic (B-) field. Since it was first proposed by Isoshima {\it et
al.} \cite{Isoshima2000}, a number of studies
\cite{Nakahara2000,Ogawa2002,Mottonen2002} have focused on this
simple protocol of creating a vortex through adiabatically flipping
the bias B-field of an Ioffe-Pritchard trap (IPT). It has been
faithfully demonstrated by several groups
\cite{Leanhardt2002,Leanhardt2003,Kumakura2006,Isoshima2007}. In
recently proposed generalizations \cite{Mottonen2007,Xu2008},
continuous vortex pumping is proposed using two sets of B-fields
with different geometries to break the otherwise time reversal
symmetry between the flip and back-flip (unflip) steps of each pump
cycle. A radial $x$-bias in the back-flip step improves the
operation efficiency, leading to a monotonically increasing
vorticity with higher and higher rotations following repeated flip
and back-flip cycles \cite{Xu2008}. While experimental challenges
remain, especially for large vorticities from repeated pump cycles,
every step of the continuous pump cycle has been demonstrated
despite various compromises in adiabaticity, heating, spin flips,
and decoherence associated with repeated controls of the B-field
gradient. What remains unclear, however, is whether the B-field
control protocols are capable of generating superposition states of
counter-circulations, reminiscent of the so-called super counter
flows \cite{Su2bec}.

In this Letter, we provide an affirmative answer that the B-field
manipulating vortex pump protocols \cite{Mottonen2007,Xu2008} remain
applicable to quantum superposition states, thus can be used to
generate counter-circulation states. Furthermore, as we shall
illustrate below through numerical simulations, the generalized
quantum pump protocols remain effective for a host of quantum
correlated states: such as single atom superpositions of
counter-circulating states, many atom superpositions of
counter-circulating states, spin squeezed, and maximally entangled
or NooN-like counter-circulating states.

Before presenting the details, we describe briefly
how an adiabatic flip of the axial $z$-bias B-field
can be used to create counter circulating states.
For an atom of hyper-fine spin $F$ in a 2D multipole B-field,
there exists a conserved quantity
$D_z=L_z-(Q-1)\times F_z$ \cite{Zhang2007,Xu2008} with
$Q=2$ ($Q=3$) for a 2D quadrupole (hexapole) field.
$\vec L$ is the OAM of the atomic center of mass motion.
An atom with no initial vorticity ($L_z=0$) and
in the $F_z=-\hbar$ state gains a circulating vorticity of
$L_z=2(Q-1)\hbar$ following an adiabatic flip of the axial bias
to the internal spin state $F_z=\hbar$. The conservation of $D_z$
also implies an atom with no initial vorticity
($L_z=0$) and in the $F_z=\hbar$ state is adiabatically
flipped into the $F_z=-\hbar$ while gaining a circulating vorticity of
$L_z=-2(Q-1)\hbar$. Both scenarios can occur simultaneously,
giving rise to a superposition of counter-circulating states,
if trapping is facilitated by an
externally applied optical trap instead of by the B-field itself \cite{Xu2008}.

For spin-1 $^{87}$Rb atoms, the two internal states $\left|\uparrow\right\rangle
\equiv|F=1, M_F=1\rangle$ and $\left|\downarrow\right\rangle\equiv|F=1,M_F=-1\rangle$
can be coupled together with two-photon Raman process, thus any
single atom superposition state of $\alpha\psi_{\uparrow}(\vec{r})\left|\uparrow\right\rangle+\beta
\psi_{\downarrow}(\vec{r})\left|\downarrow\right\rangle$ can be prepared with ease and precision,
albeit in a trap. The flipping of the axial bias then creates
a condensate state whereby every atom is in the superposition
$\alpha\psi_{\uparrow}(\vec r)e^{-2i(Q-1)\phi}\left|\downarrow\right\rangle
+ \beta\psi_{\downarrow}(\vec r)e^{2i(Q-1)\phi}\left|\uparrow\right\rangle$.
The respective mode function $\psi_{\uparrow,\downarrow}(\vec r)$ is
obtained from solving coupled
Gross-Pitaevskii equations and $\phi$ is the azimuthal angle.
Adopting the vortex pump protocol described previously \cite{Mottonen2007,Xu2008},
or as described step by step in the paragraphs to follow,
a superposition state with opposite high vorticity in each component
is prepared after $q$ pumping cycles: $\alpha\psi_{\uparrow}(\vec r)e^{-i2q(Q-1)\phi}
\left|\uparrow\right\rangle+ \beta\psi_{\downarrow}(\vec r)e^{i2q(Q-1)\phi}
\left|\downarrow\right\rangle$. This same substitution from pump
operation for
a single atom $\left|\uparrow\right\rangle\to \left|\uparrow\right\rangle
e^{-i2q(Q-1)\phi}$ and $\left|\downarrow\right\rangle\to \left|\downarrow\right\rangle
e^{i2q(Q-1)\phi}$ is transmittable to many atom states:
i.e., it can be used to generate circulation squeezing from spin
squeezed states and maximal circulation entangled state with counter
circulations
$\alpha\psi_{\uparrow\dots\uparrow}(\vec{r}_1,\dots,\vec{r}_N)
e^{-i2q(Q-1)\sum_i\phi_i}\left|\uparrow\dots\uparrow \right\rangle+
\beta\psi_{\downarrow\cdots\downarrow}(\vec{r}_1,\dots,\vec{r}_N)
e^{i2q(Q-1)\sum_i\phi_i}\left|\downarrow\cdots\downarrow
\right\rangle$ from a NooN-like state, where $\phi_i$ represents the
azimuthal angle of the $i$-th atom. A projective measurement of the
internal spin along the orthogonal direction (to
$|\downarrow\rangle$ and $|\uparrow\rangle$) interferences the two
spatial amplitudes, giving rise to fringes along the azimuthal
direction, which is sensitive to rotation as in the famous Sagnac
effect \cite{halkyard}.

\begin{figure}[H]
\centering
\includegraphics[width=2.35in]{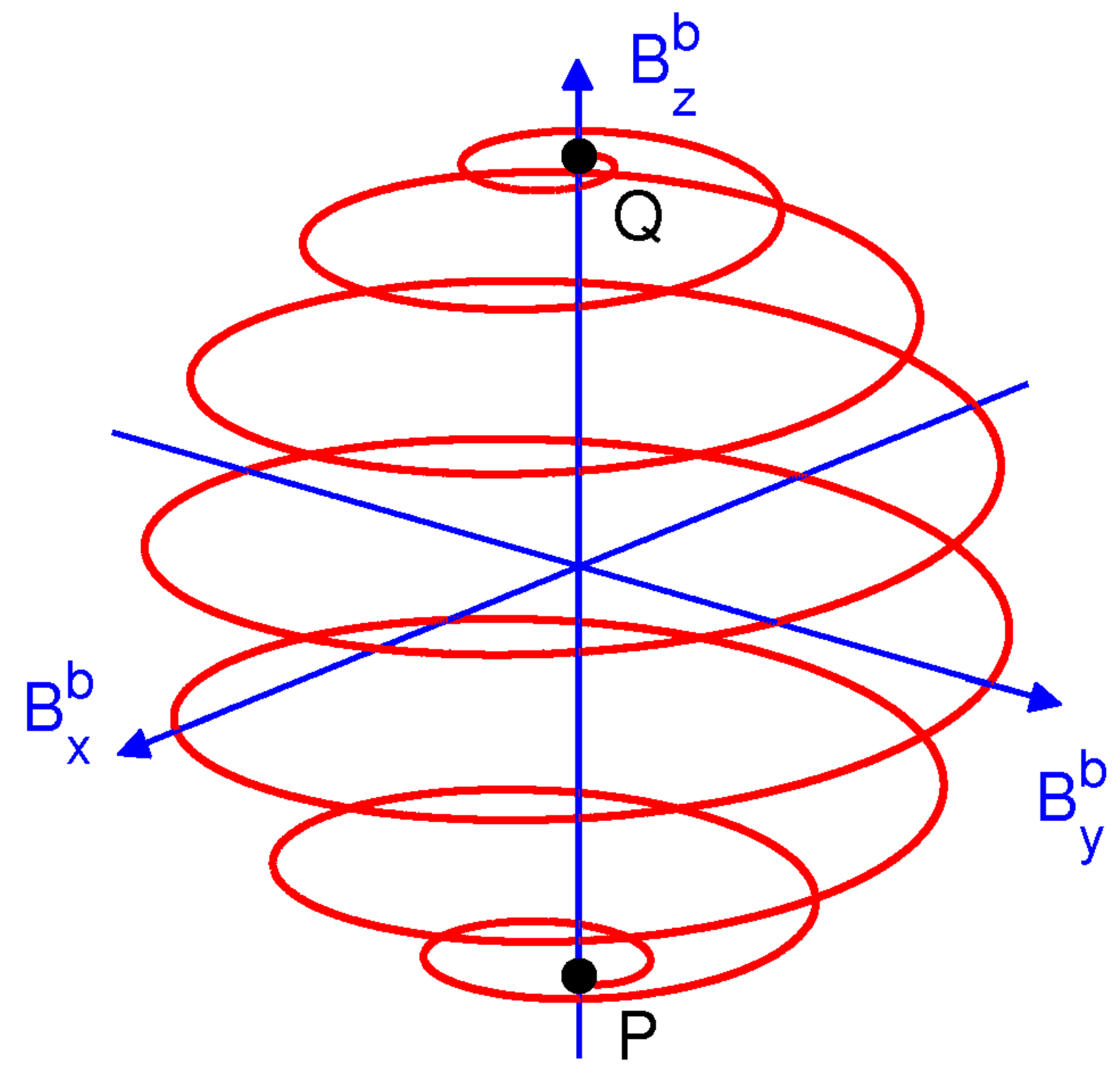}
\caption{(Color Online). The temporal trajectory
of the total bias B-field. The back-flip begins with the bias along
the $-z$-axis, depicted by the point P. In the end, it is reset to the
original direction, the point Q along the $+z$-axis, before the first step,
ready for the next pump cycle. }
\label{fig1}
\end{figure}

To facilitate experimental effort, we now discuss an improved
back-flip step for the quantum pump protocol, utilizing both bias
field controls along the transverse $x$- and $y$-axis as in a TOP
\cite{Petrich1995}. The 2D quadrupole field (2DQT) of
the IPT will be kept untouched
during the operation. In addition to the many nice features as we
illustrate below, this revised back-flip step is ideally suited
for manipulating quantum superposition states. In the first step of each
pump cycle, we still flip the bias of the IPT to create a $2F$
vortex or increase its vorticity by $2F$
\cite{Isoshima2000,Nakahara2000,Ogawa2002,Mottonen2002}. For the
second step, a previous protocol requires a radial $x$-axis bias
\cite{Zhang2007,Xu2008}. If the 2DQT were kept fixed, the zero
B-field center will move along opposite direction when the $x$-axis
bias is increased. The axial symmetry is broken, as a result, atoms
will follow to chase the low field region. The atomic spins thus
cannot be effectively rotated along the vertical direction of local
B-field, which was the reason to turn off the 2DQT before
back-flipping the atomic spin \cite{Zhang2007,Xu2008}. Based on
extensively numerical simulations and analytical investigations of
the adiabatic limits in TOP, we now find that the 2DQT can be kept
in place, provided a rotating radial bias is introduced as in a TOP
\cite{Petrich1995} when the axial $z$-bias is back-flipped. When
both $x$- and $y$-biases are timed with the axial bias
flip, we can reset the atomic spin and the IPT without
adversely affecting the vorticity already accumulated.

As a typical implementation, in the first step we adiabatically flip
the axial $z$-bias $B^b_z$ of the IPT while maintaining the 2DQT.
The time dependence of $B^b_z$ was discussed previously \cite{Xu2008}.
The initial value for $B^b_z$
is chosen to be much larger than the radial B-field so that all
atoms are properly polarized.
For a spin-1 condensate, initially all atoms are in the low-field seeking state,
the averaged spin angular momentum per atom is $F_z\simeq -\hbar$. Flipping
of the bias changes $F_z$ to $\hbar$ referenced
to the initial positive $z$-axis. The flipped internal state remains weak field
seeking because $F_z=-\hbar$ if the quantization axis were chosen to be
along the flipped B-field direction in the $-z$-axis. This
first flip is the same as before \cite{Mottonen2007,Xu2008},
which creates a vortex in a spin-1 condensate with an OAM $L_z=2\hbar$,
or increase its vorticity by $2\hbar$.
In the improved second step as outlined above,
the axial bias is back-flipped together with
a rotating radial bias as illustrated in Fig. \ref{fig1},
which are affected
if the three biases $B^b_x$, $B^b_y$, and $B^b_z$
are tuned simultaneously according to
\begin{eqnarray}
  B^b_x(t)&=&B^b_z(0)\sin\{\pi[1-(t-T_1)/T_2]\}\cos(\omega_{\rm rf}t),\nonumber\\
  B^b_y(t)&=&B^b_z(0)\sin\{\pi[1-(t-T_1)/T_2]\}\sin(\omega_{\rm rf}t),\nonumber\\
  B^b_z(t)&=&B^b_z(0)\cos\{\pi[1-(t-T_1)/T_2]\},
  \label{bias}
\end{eqnarray}
where $T_1$ and $T_2$ are the durations for the first and second steps respectively,
and $\omega_{\rm rf}$ is the frequency for the transverse bias rotation.
The transverse bias rotation
is engineered to be along $z$-axis so that the axial symmetry is
approximately maintained, and the OAM per atom $L_z$
remains conserved.
A typical B-field trap frequency is $\sim 10^2$Hz,
while the Rabi frequency between the two internal states
can be easily much larger $\sim 10^6$Hz, of the
order of the Larmor frequency. A reasonable
$\omega_{\rm rf}$ is $\sim 10^3$Hz or higher in order to assure adiabaticity as in TOP.
At the end of the back flip step, all atoms are reset to as before the
first step except for the vorticity gained. Cyclically repeating the two steps
gives rise to the new vortex pump protocol,
now with the 2DQT fixed.

Our quantum pump protocol is now demonstrated numerically for a
spin-1 $^{87}$Rb condensate with $N=10^5$ $^{87}$Rb atoms ($F=1$) in
an IPT with an optical plug
$V_p(\rho,z,\phi)=U\exp(-\rho^2/2\rho_0^2)$. The actual trapping of
atoms comes from the optical trap
$V_o=M\omega_{\perp}^2(x^2+y^2+\lambda^2z^2)/2$.
Gravity is assumed along the minus $z$-axis direction. For
simplicity, atoms are confined to the ground state $\phi(z)$
in the $z$-direction. The effective coupled 2D Gross-Pitaevskii
equations for $\psi(\vec{\rho},t)$ defined by
$\Psi(\vec{r},t)=\psi(\vec{\rho},t)\phi(z)$ becomes \cite{Xu2008a},
\begin{eqnarray}
    i\hbar\frac{\partial\psi_{\pm1}}{\partial t}&=&
    \left[\frac{}{}H_0+H_{\pm1\pm1}^{\rm ZM}+c_2^{(\rm 2D)}(n_{\pm1}+n_0-n_{\mp1})\right]\psi_{\pm1}\nonumber\\
    &&+c_2^{(\rm 2D)}\psi_{\mp1}^*\psi_0^2+H^{\rm ZM}_{\pm10}\psi_0+H^{\rm ZM}_{\pm1\mp1}\psi_{\mp1},\nonumber\\
    i\hbar\frac{\partial\psi_0}{\partial
    t}&=&\left[\frac{}{}H_0+H_{00}^{\rm ZM}+c_2^{(\rm 2D)}(n_1+n_{-1})\right]\psi_0\nonumber\\
    &&+2c_2^{(\rm 2D)}\psi_0^*\psi_1\psi_{-1}
    +H^{\rm ZM}_{01}\psi_1+H^{\rm ZM}_{0-1}\psi_{-1}.
    \label{gpe}
\end{eqnarray}

For the optical traps, we take $\lambda=50$ and
$\omega_{\perp}=2\pi\times 60\, \rm Hz$, $U/\hbar=2\times10^5\, \rm
Hz$, and $\rho_0=5\, \rm \mu m$. The total B-field is
$\vec{B}=B'(x\hat{x}-y\hat{y})+B_x^b\hat{x}+B_y^b\hat{y}+B_z^b\hat{z}$,
where the first term is the familiar 2DQT, which is fixed during the
whole process, and is responsible for the geometrical phase
structure of the vortex state. $B'$ is chosen to be $5\, \rm G/cm$. In the
first step of each cycle, biases $B_x^b$ and $B_y^b$
are turned off, and the $z$-axis bias $B_z^b$ is flipped
according to the time dependence of Eq. (4) of \cite{Xu2008} with
$B_z^b(0)=0.1\, \rm G$. In the second step, all three biases
are programmed according to the time dependence of
Eq. (\ref{bias}), with $T_1=45\, \rm ms$, $T_2=15\,
\rm ms$, and $\omega_{\rm rf}=2\pi\times5\, \rm kHz$.

The initial superposition state is prepared by imaginary time
propagation of coupled Gross-Pitaevskii equations (\ref{gpe})
starting from
$\psi_G(\left|\uparrow\right\rangle_B+\left|\downarrow\right\rangle_B)$,
with $\psi_G$ a Gaussian wave function and the subscript $B$
denotes local B-field quantization, under a fixed
IPT $B'(x\hat{x}-y\hat{y})+B_z^b(0)$.
The two states $\left|\uparrow\right\rangle_B$ and $\left|\downarrow\right\rangle_B$
are essentially the same as $\left|\uparrow\right\rangle$ and $\left|\downarrow\right\rangle$
whenever the B-field bias $B_z^b(0)$ dominates.
At the end of the imaginary time propagation, the state takes
approximately the form $\left(\psi_{\uparrow}\left|\uparrow\right\rangle
+\psi_{\downarrow}\left|\downarrow\right\rangle\right)/{\sqrt{2}}$,
from which we then apply our quantum pump protocol in real time.
In the first step of each cycle, the $z$-axis bias $B_z^b$ is flipped from
$0.1\, \rm G$ to $-0.1\, \rm G$, in $45\, \rm ms$.
A vorticity of $2\hbar$ ($-2\hbar$) is pumped into
the component $\left|\uparrow\right\rangle$ ($\left|\downarrow\right\rangle$).
The second step keeps the vorticity,
resetting the axial bias and the atomic spins.

\begin{figure}[H]
\centering
\includegraphics[width=3.3in]{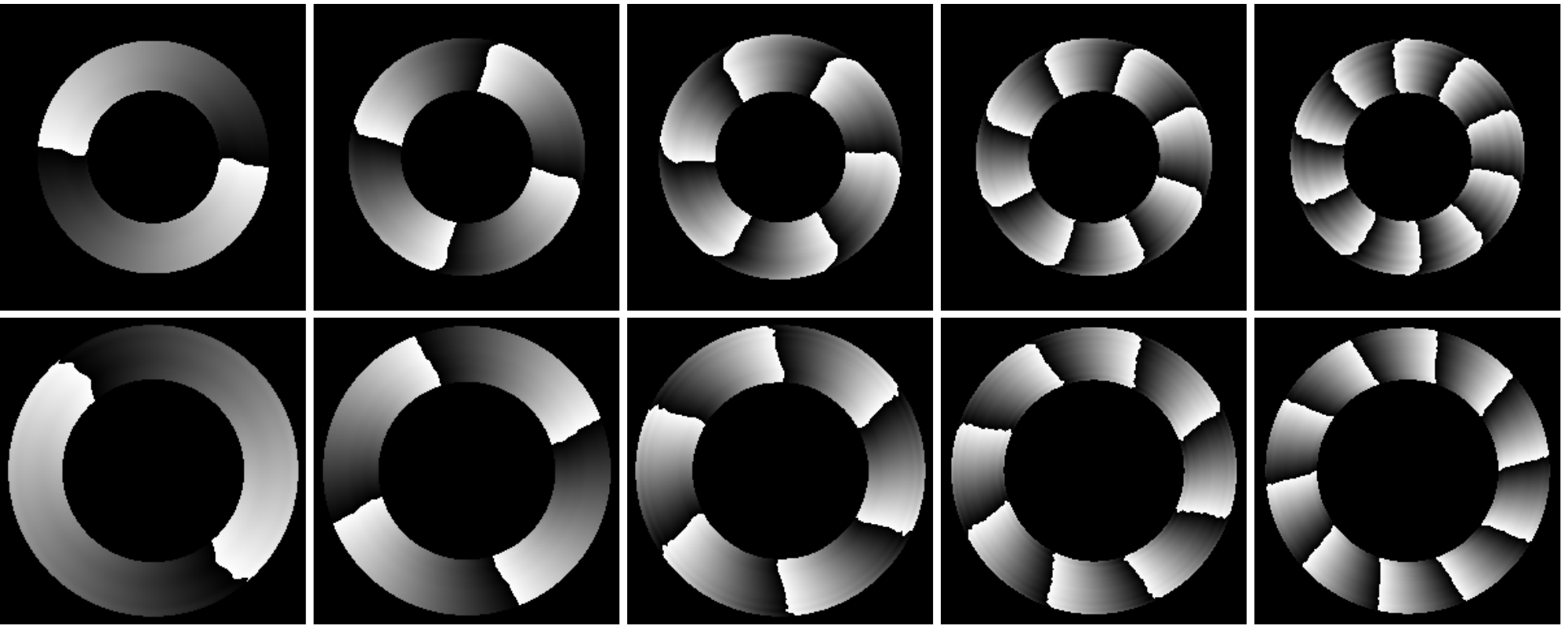}
\caption{The temporal development of the spatial
phase structures from the quantum pump
with a rotating transverse bias
$\vec{B}_h=B'(x\hat{x}-y\hat{y})+B_x^{b}\hat{x}+B_y^{b}\hat{y}+B_z^{b}\hat{z}$.
The first(second) row shows the $\left|\uparrow\right\rangle$
($\left|\downarrow\right\rangle$) component at
the end of the first flips, or in the middle of each cycle at $45$,
$105$, $165$, $225$, and $285$ ms.
The maximum for $B_x^{b}$, $B_y^{b}$, and $B_z^{b}$ are
$0.1\, \rm G$, $0.1\, \rm G$, and $0.1\, \rm G$,
respectively. White (black) color denotes phase $-\pi$ ($\pi$). }
\label{fig2}
\end{figure}

Figure \ref{fig2} illustrates the intended time development
of the spatial phase structure.
The first (second) row shows the $\left|\uparrow\right\rangle$
($\left|\downarrow\right\rangle$) component at
the end of the first flips, or in the middle of each pump cycle at
various times respectively.
Despite the presence of a 2DQT,
the vortex pump works nicely as predicted. It increases and decreases respectively
circulations for the initial $\left|\downarrow\right\rangle$ and $\left|\uparrow\right\rangle$
components, or it enlarges their
respective vorticities during the first step of each pump cycle.
The vorticities gained are maintained during the
second steps of each cycle while the system resets. The whole
pump scheme is now considerably simpler, and
from the experimental point of view, can be implemented with
mature technologies,
particularly at laboratories with TOPs already in place.

%

The above discussion and numerical demonstration concerns
the quantum pump protocol applied to a condensate of
all atoms in the same two state superposition
$\left[\psi_{\uparrow}(\vec{r})\left|\uparrow\right\rangle
+\psi_{\downarrow}(\vec{r})\left|\downarrow\right\rangle\right]^{\otimes N}/2^{N/2}$.
As was alluded to earlier, our quantum pump
also applies to more general superposition states,
e.g., spin squeezed states and
 maximally entangled N-GHZ or NooN state.
Because of the limitations in simulating
spatial wave functions for a many atom state,
as an illustration, we consider the simple case of an
entangled two atom
state $\left[\psi_{\uparrow\uparrow}(\vec{r}_1,\vec{r}_2)
\left|\uparrow\uparrow\right\rangle+\psi_{\downarrow\downarrow}(\vec{r}_1,\vec{r}_2)
\left|\downarrow\downarrow\right\rangle\right]/\sqrt{2}$,
which provides the crucial link to generalize our pump
protocol to many atom systems.
After the first step of bias flip, the initial state
evolves into $\left[\psi_{\uparrow\uparrow}
e^{-i2(\phi_1+\phi_2)}\left|\downarrow\downarrow\right\rangle+\psi_{\downarrow\downarrow}
e^{i2(\phi_1+\phi_2)}\left|\uparrow\uparrow\right\rangle\right]{\sqrt{2}}$
according to our quantum pump protocol. It further evolves into
$\left[\psi_{\uparrow\uparrow}
e^{-i2(\phi_1+\phi_2)}\left|\uparrow\uparrow\right\rangle+\psi_{\downarrow\downarrow}
e^{i2(\phi_1+\phi_2)}\left|\downarrow\downarrow\right\rangle\right]{\sqrt{2}}$
during the second step of back-flip,
with the same spatial phase structure intact as at the end of the first pumping cycle.
The numerical simulations use the same parameters
as in Fig. \ref{fig2}, but with $\omega_{\perp}=2\pi\times 100\, \rm Hz$,
$\lambda=30$, and $\rho_0=2\, \rm\mu m$. For simplicity, we choose to
neglect interactions between the two spin-1 $^{87}$Rb atoms.

\begin{figure}[H]
\centering
\includegraphics[width=3.2in]{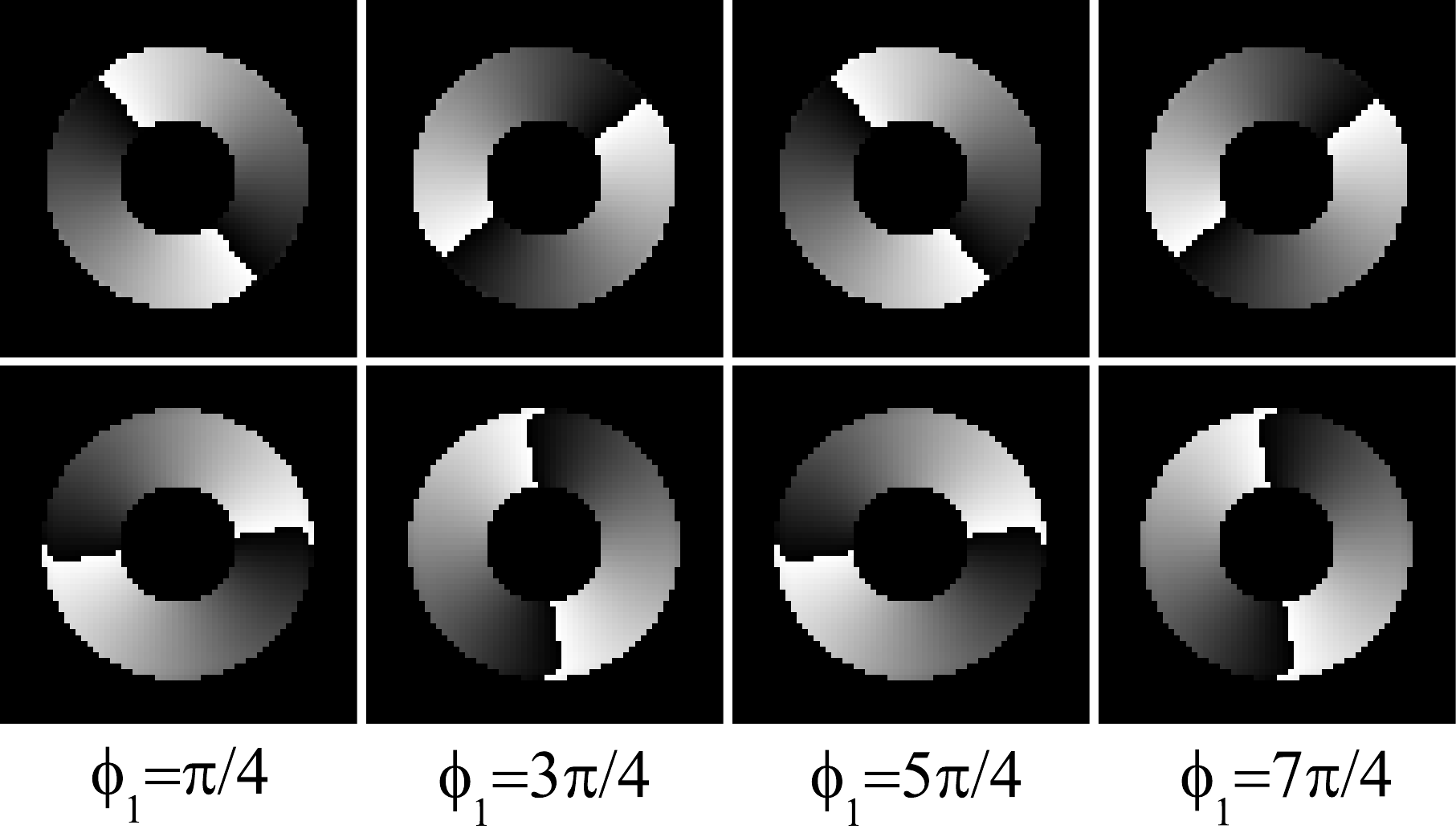}
\caption{The vortical phase structures for two atoms
with one fixed at azimuthal angle
$\phi_1=\pi/4, 3\pi/4, 5\pi/4$, and $7\pi/4$ from the left
to the right column. The first (second) row shows
the spatial distribution for the remaining phase
of the wave function component $\left|\uparrow\uparrow\right\rangle$
($\left|\downarrow\downarrow\right\rangle$ after the first step of bias flip.
 White (black) color denotes phase $-\pi$ ($\pi$). }
\label{fig3}
\end{figure}

Figure \ref{fig3} demonstrates the actual temporal
development for
the phase structures of the two atom wave function
$\psi_{\sigma_1\sigma_2}(\vec{r}_1,\vec{r}_2)$. To simplify the
graph, we fix one atom at the radius corresponding to the maximum
density $\sum_{\sigma_1\sigma_2}\int
d\vec{r}_2|\psi_{\sigma_1\sigma_2}(\vec{r}_1,\vec{r}_2)|^2$, and
specific azimuthal angles $\phi_1$ where the spin state for one atom
is $M_F=\sigma_1$ while $M_F=\sigma_2$ for the other. The first
(second) row reveals perfect vortical structure for the the
$\left|\uparrow\uparrow\right\rangle$
($\left|\downarrow\downarrow\right\rangle$ components after the
first step of bias flip. Similar results are obtained for the
$\left|\downarrow\downarrow\right\rangle$
($\left|\uparrow\uparrow\right\rangle$ components at the end of each
pumping cycle. We thus show for two atomic states
$\left|\uparrow\uparrow\right\rangle$
($\left|\downarrow\downarrow\right\rangle$), each one gains a
vorticity $-2\hbar$ ($2\hbar$) irrespective how the other atom
evolves. The vorticity gained is further shown to remain preserved
after the back-flip of the bias. Since the two atoms we adopt are
identical, the vorticity gained in the two atoms will be the same
during the whole protocol, consistent with our previous discussion.
Not surprisingly, these results can still be understood by
considering the conservation of $D_z=L_z-F_z$.

The two atomic internal states we illustrate here are very
different; state $|F=1,M_F=-1\rangle$ is low-field seeking, while
$|F=1,M_F=1\rangle$ is high-field seeking. More generally, any two
states with opposite $M_F$ can be used as long as they can both be
confined within the geometrical pole region of a 2DQT. When the two
states are of similar magnetic properties, like the two week field
seeking states $|F=2,M_F=1\rangle$ and $|F=1,M_F=-1\rangle$ of
$^{87}$Rb atoms \cite{Myatt1997}, additional optical traps become
less relevant. Moreover, our protocol is checked to work equally
well for superpositions with different amplitudes $\alpha$ and
$\beta$, and for superfluid mixtures of different atomic species
\cite{Bargi2007}. The validity for our pump protocol
can be confirmed through interfering the two
amplitudes \cite{Kapale2005,Wright2008}.

In conclusion, we propose a practical and effective
quantum vortex pump scheme capable of increasing/decreasing
vortical circulations of
an arbitrary superposition for two state atoms.
An immediate application concerns the generation of
superposition states exhibiting
nonclassical circulation properties such as from
spin squeezed states
with enhanced signal to noise ratios upon on suitable measurement.
Our work maps enhanced sensing capabilities of correlated
many atom states to rotational sensing,
and shines new light on a number of interesting topics
from sensitive interferometers to test of
rotational equivalence principle \cite{zhangyuanzhong}.

This work is supported by NSF of China under Grant No. 10640420151,
and NKBRSF of China under Grants No. 2006CB921206 and No.
2006AA06Z104.

\end{document}